\documentclass[superscriptaddress,aps,preprint]{revtex4}
\usepackage{graphicx}

\begin{document}

\title{Structural motifs of biomolecules}

\author{Jayanth R. Banavar}
\affiliation{Department of Physics, 104 Davey Lab,
The Pennsylvania State University, University Park PA 16802, USA}
\author{Trinh Xuan Hoang}
\affiliation{Institute of Physics and Electronics, VAST,
10 Dao Tan, Hanoi, Vietnam}
\author{John H. Maddocks}
\affiliation{Laboratory for Computation and Visualization in
Mathematics and Mechanics, EPFL FSB IMB, Ecole Polytechnique
Federale de Lausanne CH-1015 Lausanne, Switzerland}
\author{Amos Maritan}
\affiliation{Dipartimento di Fisica `G. Galilei',
Universit\`a di Padova, Via Marzolo 8, 35131 Padova, Italy}
\affiliation{CNISM, Unit\`a di Padova, Padua, Italy}
\affiliation{INFN, Sezione di Padova, Padua, Italy} \author{Chiara
Poletto} \affiliation{Dipartimento di Fisica `G. Galilei',
Universit\`a di Padova, Via Marzolo 8, 35131 Padova, Italy}
\author{Andrzej Stasiak} \affiliation{Faculty of Biology and Medicine,
Center for Integrative Genomics, University of Lausanne, Lausanne CH
1015, Switzerland} \author{Antonio Trovato} \affiliation{Dipartimento
di Fisica `G. Galilei', Universit\`a di Padova, Via Marzolo 8, 35131
Padova, Italy}
\affiliation{CNISM, Unit\`a di Padova, Padua, Italy}

\begin{abstract}
Biomolecular structures are assemblies of emergent anisotropic
building blocks, uniaxial helices and biaxial strands. We provide a
conceptually novel approach to understand a marginally compact phase
of matter that is occupied by proteins and DNA. This phase, that is in
some respects analogous to the liquid crystal phase for chain
molecules, stabilizes a range of shapes that can be obtained by
sequence independent interactions occurring intra- and
inter-molecularly between polymeric molecules. We present a
singularity free self-interaction for a tube in the continuum limit
and show that this results in the tube being positioned in the
marginally compact phase. Our work provides a unified framework for
understanding the building blocks of biomolecules.
\end{abstract}

\maketitle

\section{Introduction}

The structures and phases adopted by inanimate matter have
traditionally been understood and predicted by simple paradigms,
e.g. seemingly disparate phenomena such as phases of matter,
magnetism, critical phenomena, and neural
networks~\cite{BialekNature} have been successfully studied within
the unified framework of an Ising model~\cite{hardsphere}.  Liquid
crystals~\cite{DeGennes}, whose molecules are not spherical, form
several distinct stable, yet sensitive structures. They possess
translational order in fewer than three dimensions and/or
orientational order and exist in a phase between a liquid with no
translational order and a crystal with translational order in all
three directions. The liquid crystal phase is poised in the
vicinity of the transition to the liquid phase and accounts for
its exquisite sensitivity. Any material that resides in a
particular phase of matter exhibits the general properties
characteristic of that specific phase and there are just a few
essential ingredients, such as the symmetry of the atoms or
molecules comprising the material and certain macroscopic
parameters such as the pressure and
temperature, that determine the relevant phase.

Biomolecules, such as DNA and proteins, form the basis of life and
exhibit simple forms such as a single, double or a triple helix and
almost planar sheets assembled from zig-zag strands~\cite{tbmb}. The
latter are also implicated in amyloid structure which play a role in
diseases such as Alzheimer's and Type II
diabetes~\cite{ChitiDobson}. The origin of these structures is now
well understood based on details of the constituent atoms and the
quantum chemistry that governs their
assembly~\cite{Pauling,Pauling1,Pauling2}. The common use of these
modular structures by nature begs for a simple unified explanation for
their ubiquity. Here we show that not only single, double and triple
helices but also planar sheets made up of biaxial strands are natural
forms in a conceptually novel, marginally compact phase of matter of a
flexible tube, the simplest description of a chain molecule that
incorporates the correct symmetry. Remarkably, this phase of matter is
analogous to a liquid crystalline phase but for chain molecules and is
assembled from emergent anisotropic building blocks. Our work provides
a unified description, which transcends chemical details, of the
structural motifs of biomolecules. We elucidate the role played by
discreteness in promoting the creation of biaxial strands through
spontaneous symmetry breaking. An important consequence of our work is
that it suggests that physical scientists and engineers who wish to
build nifty machines akin to proteins would do well to design their
devices so that they are poised in this phase of matter with all its
advantages.

The fluid and crystalline phases of ordinary matter are well
described by a simple model of a collection of beads or hard
spheres~\cite{hardsphere}. A hard sphere can be thought of as a
point, a zero dimensional object, in space with an excluded volume
region obtained by symmetrically inflating it to a size equal to
its radius.  The packing of spheres is a classic optimization
problem \cite{Szpiro} with a long and venerable history and many
important applications. There are a large number of important
synthetic materials, such as plastic, rubber, gels, and textile
fibres, comprised of polymer molecules~\cite{DeGennes2}. Life is
also based on chain molecules such as DNA and proteins. The
generalization of the hard sphere to a one dimensional manifold
consists of taking a curve and symmetrically inflating it to form
a flexible tube of thickness $\Delta$ characterized by {\em
uniaxial symmetry} (Figure 1). We will show that the tube paradigm
provides a unified and natural explanation for helical forms and
sheets in biomolecules.

\section{Results and discussion}

\subsection*{Buried area of tubes}

We begin by describing the discrete version of a tube of thickness
$\Delta$ represented by a chain of coins, whose planes coincide with
equally spaced circular cross-sections of the tube. The
self-avoidance of the tube is implemented by the three-body
prescription~\cite{GM,JSP} described in the caption of Figure 1. A
classic way to take into account the solvophobic effect is to
introduce an attractive pairwise interaction between the coin
centers with an interaction range $R_0$. In \cite{Davide1,Davide2},
it was shown that, when $\Delta \sim R_0$, a short tube had
relatively few low energy conformations compared to the generic
compact phase ($\Delta \ll R_0$) and the swollen phase ($\Delta \gg
R_0$). Structures in this novel marginally compact phase were found
to be constructed from two building blocks, helices and zig-zag
strands, and able to possess liquid crystal like sensitivity because
of being poised in the vicinity of a transition to the swollen
phase.

In the continuum limit, two-body interactions are necessarily
singular because there is a continuum of pairs, close by along the
chain, that are within the interaction range \cite{JSP}. A
singularity-free formulation of the attractive interaction follows
from the following physical situation. Let us suppose that the tube
is immersed in a poor solvent whose molecules are approximated by
spherical balls of radius $R$. For any given tube configuration,
there are regions of the tube surface that the solvent molecules can
come in contact with and other regions that are unaccessible. The
latter constitutes the buried surface of the tube configuration. The
bigger the radius of the solvent molecule radius, the larger is the
buried surface. We will show that an interaction based on the buried
area is sufficient to lead to ground state tube conformations in the
marginally compact phase with a variety of secondary motifs.

The simplest potential for the solvophobic interaction is given by
\begin{equation}\label{H}
\mathcal{H}_0(\{{\bf r}(s)\}) = -\kappa_{sol}\Sigma_B (R),
\end{equation}
where ${\bf r}(s)$ defines the smooth curve corresponding to the tube
axis, $s$ is the arc-length, $\Sigma_B(R)$ is the buried area in
presence of solvent molecules of radius $R$ and $\kappa_{sol}$ is an
effective interaction strength which we set equal to one without loss
of generality. An analytical derivation is given in Methods.

The interaction as given by Eq.(\ref{H}) describes a tube with a
uniform solvophobicity. As discussed below, for proteins it is more
appropriate to introduce a mixed solvophobicity tube. In its simplest
version this tube has two types of surface regions, each characterized
by a different degree of solvophobicity, as described by
Eq. (\ref{MS}) in Methods.

\subsection*{Uniform solvophobic tubes}

Figure 2 (a-h) depicts the conformations adopted by short tubes
subject to compaction. Our results are obtained by maximizing the
buried surface area \cite{surface1,surface2,surface3} of the tube
(see eqs.(\ref{H}) and (\ref{BA})). Such an optimization
requirement is generically encountered when a tube shows a higher
affinity to itself than to a solvent, e.g.  in poor solvent
conditions~\cite{DeGennes2}. Strikingly, the conformations of
choice are single, double and triple helices, all characterized by
chirality and adopted by nature in the context of biomolecules
such as proteins and DNA. It is remarkable that the shapes of
close packed single and double helices adopted by flexible tubes
match those of $\alpha$-helices~\cite{Tubone,KamienScience} and
the DNA double-helix~\cite{stasiak}, respectively. At somewhat
higher temperatures, one obtains the conformation in Figure 2 (h),
as a result of the interplay between the simultaneous maximization
of both the buried area and the entropy, comprising almost
parallel elongated tube segments. Figure 3 shows an optimal
arrangement of several segments of continuous tubes arranged in a
hexagonal array. (Such an arrangement is called an Abrikosov
flux-lattice in the field of superconductivity~\cite{Abrikosov}.)
One would expect that the helical state of a few tubes would be
supplanted by such an
ordering when many tubes are packed together.

\subsection*{Mixed solvophobic tubes}

We turn now to a consideration of two distinct mechanisms that
promote sheet formation within the context of the tube picture. The
first mechanism is directly inspired from the observation that the
side-chains of amino acids stick out in a direction approximately
opposite to the bending direction of the protein
backbone~\cite{Levitt} yielding an effectively mixed solvophobic
tube.  In other words, certain parts of the solvophobic tube,
determined by the instantaneous tube conformation, are already
protected from the solvent by the side chains whereas the rest of
the tube needs to shield itself from the solvent by means of the
compaction process.  The structures (i-l) in Figure 2, obtained by
minimizing the energy given by Eq.(\ref{MS}), are optimal
conformations for a single tube (i) and for multiple tubes at low
temperature (j) and at a higher temperature (k,l). Our studies have
been carried out at $c=5$ which corresponds to the region $P$ being
solvophilic and the region $H$ being solvophobic. The resulting
sheet structure is characterized by planarity as well as strands
that zigzag
normal to the plane of the sheet, as observed in real protein structures.

The second mechanism is more subtle and does not invoke mixed
solvophobicity but instead arises from the consideration of a
discrete version of the tube as described above. Instead of
maximizing the buried surface area of the continuous tube, we now
seek to maximize the number of pairwise contacts between
non-consecutive coin centers within a prescribed mutual distance of
the order of the tube thickness in order to be within the marginally
compact phase \cite{Davide1,Davide2}. The optimal packing for the
discrete case at its edge of compaction is shown in Figure 4 -- one
obtains a planar arrangement of chains which zigzag within the
plane. In the continuum, one retains the uniaxial anisotropy
characteristic of a tube whereas in the discrete case, the symmetry
between the two directions perpendicular to the principal strand
direction is {\em spontaneously broken} (in the mixed solvophobic
tube, this symmetry is broken overtly). Strikingly, the out-of-plane
zigzag pattern shown in Figure 2 for continuous tubes is realized in
protein $\beta$-sheets when viewed in the $C^{\alpha}$
representation (see Figure 5a), whereas the in-plane zigzag pattern
of the discrete case shown in Figure 4 is obtained in a different
representation with interaction centers on
both the $N-H$ and $C-O$ bonds (see Figure 5b).

\subsection*{Conclusions}

Our work suggests that protein native state structures occupy a
novel phase of matter corresponding to that of compact
conformations of a flexible tube. This marginally compact phase is
analogous in several respects to the liquid crystal phase but this
time for chain molecules. The liquid crystal phase is exquisitely
sensitive to perturbations because it is poised close to the
transition to the liquid phase. Likewise, protein structures are
able to facilitate the various range of functions that proteins
perform in the living cell because a tube at the edge of
compaction is in the vicinity of a swollen phase in which the
attractive potential is no longer operational. And just as liquid
crystals are made up of anisotropic molecules, protein native
state structures are made up of emergent anisotropic building
blocks -- uniaxial helices and almost planar sheets comprised of
biaxial strands. It is tempting to speculate that nature, through
evolution and natural selection, has been able to exploit the
marginally compact phase of flexible tubes under the constraints
of quantum chemistry governing covalent and hydrogen
bonds.

\section{Methods}

We derive an analytical expression for the buried area of a tube of
length $L$ ($0\ \le\ s\ \le\ L$) and radius $\Delta$. A generic point
on the tube surface is given by
\begin{equation}
{\bf u} (s,\theta) = {\bf r}(s) + \Delta (\hat{{\bf n}} (s) \cos
\theta + \hat{{\bf b}} (s) \sin \theta),
\end{equation}
where $\hat{{\bf n}} (s)$ and $\hat{{\bf b}} (s)$ are the normal
and binormal vectors at position $s$ respectively, and $\theta$ is
an azimuthal angle running from 0 to $2\pi$ \cite{Dubrovin}. The
surface element is given by $\sqrt{g(s,\theta)}\,ds d\theta$ ,
where
\begin{equation}
g \equiv \left|
\begin{array}{cc}
\left(\frac{\partial {\bf u}}{\partial s} \right)^2 &
\frac{\partial {\bf u}}{\partial s}\frac{\partial {\bf
u}}{\partial
\theta} \\
\frac{\partial {\bf u}}{\partial \theta}\frac{\partial {\bf
u}}{\partial s} & \left(\frac{\partial {\bf u}}{\partial \theta}
\right)^2
\end{array}
\right|.
\end{equation}

Note that $\partial^2 \hat{{\bf r}}/\partial s^2 = \kappa \hat{{\bf
n}}(s)$; $R_c(s) \equiv 1/\kappa(s)$ is the local radius of
curvature; and $\partial \hat{{\bf b}}/\partial s = -\tau \hat{{\bf
n}}(s)$, where $\tau$ is the torsion~\cite{Kamien}. One obtains
\begin{equation}
g = \Delta^2 ( 1-\Delta \kappa \cos\theta)^2.
\end{equation}
For the tube $R_c(s) = 1/\kappa(s) \ge \Delta$, $\forall s$, thus $
1-\Delta \kappa \cos\theta \ge 0$, and
\begin{equation}
\sqrt{g} = \Delta ( 1-\Delta \kappa \cos\theta).
\end{equation}
The total area of the tube is
\begin{equation}
\Sigma = \int_0^{2\pi} d\theta \int_0^L ds \sqrt{g} = 2\pi \Delta
L.
\end{equation}
The buried surface is determined by the inequality
\begin{equation}
B_R(s,\theta) \equiv \min_{s'} | {\bf r}(s) + (\Delta + R)
(\hat{{\bf n}} (s) \cos \theta + \hat{{\bf b}} (s) \sin \theta)-
{\bf r} (s') | < \Delta + R ,
\end{equation}
yielding an expression for the buried area:
\begin{equation}\label{BA}
\Sigma_B (R) = \Delta \int_0^L ds \int_0^{2\pi} d\theta
\left(1-\frac{\Delta}{R_c(s)} \cos \theta \right) \cdot
\Theta[\Delta+R - B_R(s,\theta)],
\end{equation}
where $\Theta(x)$ is equal to 1 if $x>0$ and 0 otherwise.

The simplest version of a mixed solvophobicity tube has two types of
surface regions, each characterized by a different degree of
solvophobicity. We denote these regions as $P$ (for solvophilic) and
$H$ (for solvophobic) respectively. We consider the generalized
Hamiltonian
\begin{eqnarray}\label{MS}
\mathcal{H}_c(\{{\bf r}(s)\}) & = & -\Delta\int_0^L  ds
\int_0^{2\pi}d\theta \left(1 -
\frac{\Delta}{R_c(s)}cos\theta\right) \cdot \{\Theta[\Delta + R
-B_R(s,\theta)] -
\\ & & -c\ \Theta[\theta_1 - |\pi-\theta|] \nonumber
\cdot (1-\Theta[\Delta + R -B_R(s,\theta)])\}  ,
\end{eqnarray}
where $\theta_1$ defines half the angular width of region $P$
centered around $\theta = \pi$ and $c$ is a measure of the
coupling between this region and the solvent. The case $c=0$
corresponds to the uniform tube described previously.

\begin{acknowledgments}

This work was supported by PRIN no. 2005027330 in 2005, INFN, the NSC
of Vietnam, and Swiss National Foundation grant 3100A0-11627.

\end{acknowledgments}

\begin{figure}[h]
\includegraphics[height=10.0cm]{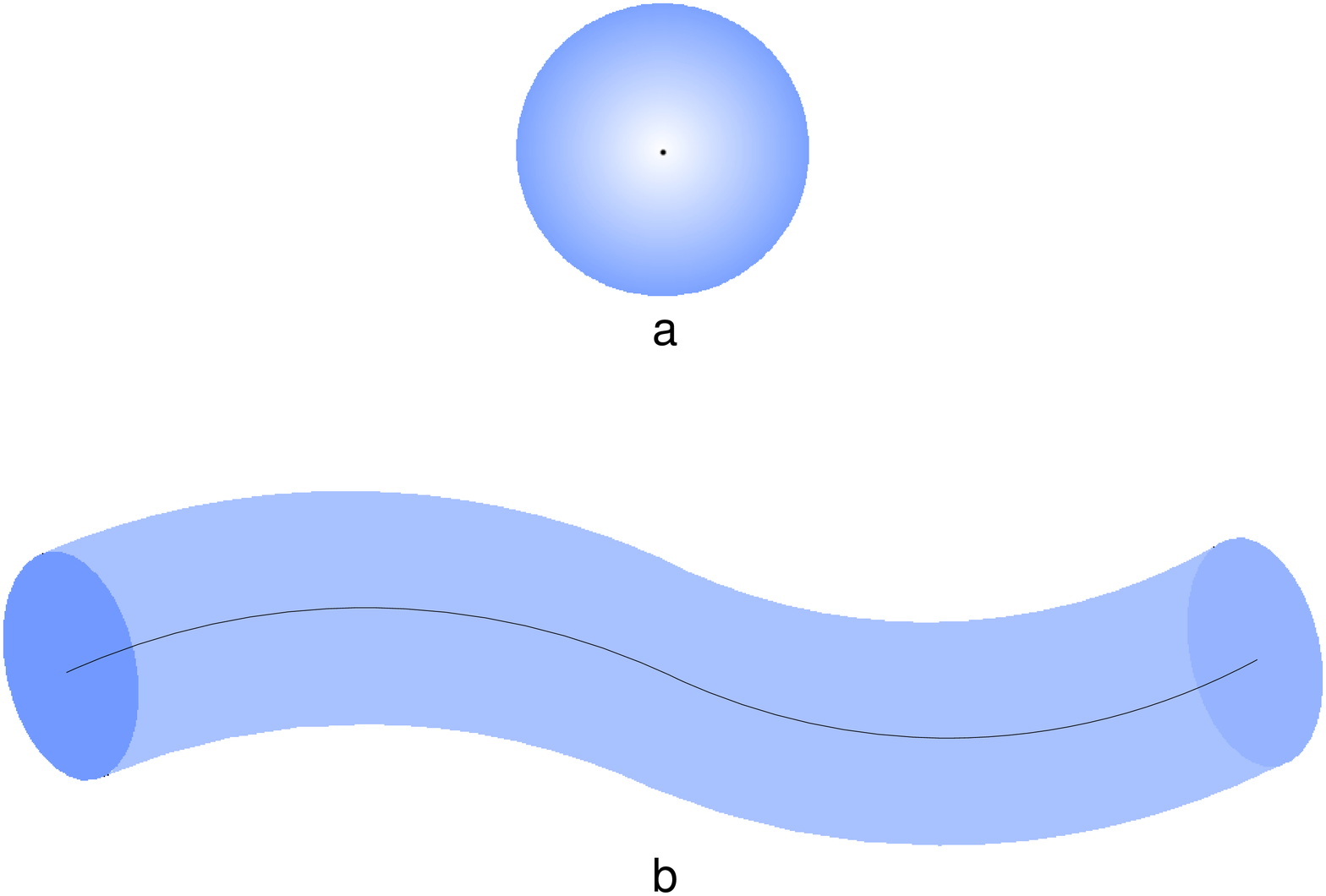}
\caption{Sketch of a hard sphere and a tube. The self-avoidance
of an ensemble of hard spheres, each of radius $\Delta$, can be
ensured by considering all pairs of spheres and requiring that none of
the distances between the sphere centers is less than $2\Delta$. The
self-avoidance of a tube of thickness $\Delta$ can be enforced through
a suitable three-body potential~\cite{GM,JSP}. We denote the tube axis
by a smooth curve, ${\bf r}(s)$, where the arc-length $s$ satisfies
$0\le s \le L$ and $L$ is the total length of the tube.  For a tube
one considers {\em all} triplets of points ${\bf r}_i={\bf r}(s_i),
i=1,2,3$ on the tube axis and draws circles through them and requires
that none of the radii $r({\bf r}_1,{\bf r}_2,{\bf r}_3)$ of these
circles is less than the tube radius $\Delta$.}
\end{figure}

\newpage

\begin{figure}[h]
\includegraphics[height=9.0cm]{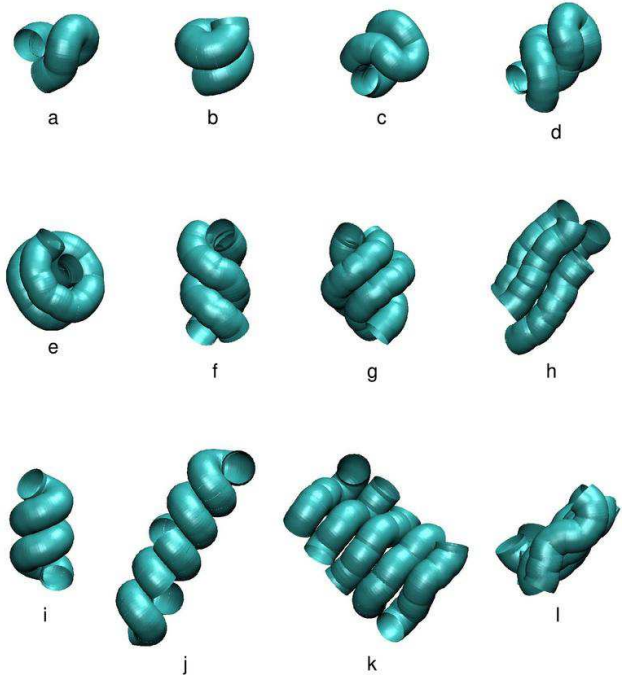}
\caption{Optimal conformations of tubes subject to compaction. In our
simulations, we considered a discretized representation with $N$
segments separated by a distance $b=\Delta/2$, where $\Delta$ is the
tube radius. The continuum limit is obtained when $b << \Delta$ -- we
have verified that our results are substantially the same on reducing
the value of $b$ down to $\Delta/3$. The conformations are obtained
using Metropolis Monte Carlo simulations by annealing or by long
simulations at constant temperature. The simulations are performed
with standard pivot and crank-shaft move sets~\cite{Sokal}. For
systems of multiple chains, the tubes are placed inside a hard-wall
cubic box of side of 40$\Delta$ -- we have verified that the walls of
the box do not influence the conformations shown. (a-e) Conformations
of single solvophobic tubes ($c=0$) of various lengths and for
different solvent molecule radius $R$ that maximize the buried area
(Eq. 8): a) $N=20$ and $R=0.1\Delta$, b) $N=20$ and $R=0.2\Delta$, c)
$N=30$ and $R=0.1\Delta$, d) $N=40$ and $R=0.1\Delta$, e) $N=50$ and
$R=0.2\Delta$; (f-h) Optimal conformations of multiple solvophobic
tubes ($c=0$) of length $N=20$ and for $R=0.1\Delta$ obtained in long
simulations at constant temperatures: f) two tubes at a low
temperature, g) three tubes at a low temperature, h) four tubes at an
intermediate temperature; (i-l) Conformations of mixed solvophobicity
tubes ($c=5$) that minimize the energy (Eq. 9).  $R=\Delta/2$ for all
cases.  i) A single helix of length $N=30$ and $\theta_1=15^o$
obtained by slow annealing (one obtains the same conformation for
$\theta_1 = 30^o$ or $45^o$). j) A stack of 4 helices of length $N=15$
and $\theta_1 = 45^o$ obtained by slow annealing.  k-l) Two views of a
planar sheet arrangement of 5 chains of length $N=15$ and $\theta_1 =
30^o$ obtained in a constant temperature simulation run at $T=0.4$.}
\end{figure}

\newpage

\begin{figure}[h]
\includegraphics[height=10.0cm]{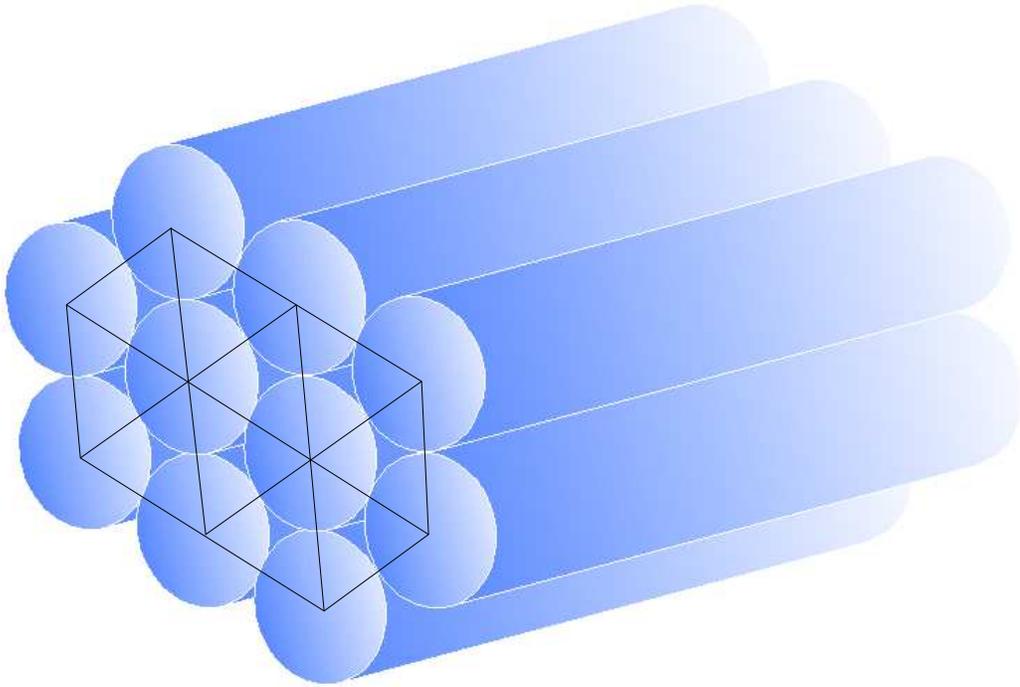}
\caption{ Optimal arrangement of several segments of
continuous tubes arranged in an Abrikosov flux-lattice like
state~\cite{Abrikosov} with straight tube segments parallel to each
other. In the plane orthogonal to the tube axes, the tube cross
sections are arranged in a hexagonal array.}
\end{figure}

\newpage

\begin{figure}[h]
   \begin{center}
     \includegraphics*[height=10.0cm]{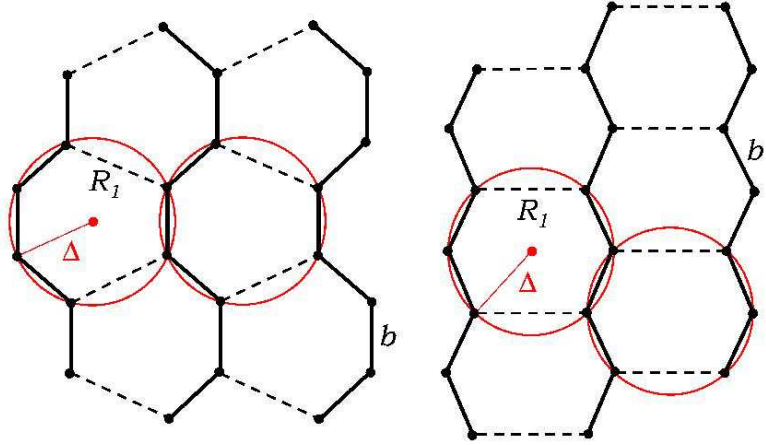}
   \end{center}
\caption{Sketch of optimally packed short segments of three
tubes (dark lines) obtained from Metropolis Monte Carlo annealing
simulations at the edge of compaction. The self-avoidance of a
discrete tube (defined through a set of $N$ points along the
discretized tube axis $\{ {\bf r}_1, {\bf r}_2, \ldots, {\bf r}_N \}$
with unit spacing ($b=1$) between consecutive points) is enforced
through the three-body potential defined in the caption of Figure
1. The number of pairwise contacts between non-consecutive beads is
maximized. Any two such beads are forbidden to come closer than 1.1
units and are defined to form a contact when they come closer than 1.6
units ($R_1=1.6$) (these numbers have been selected to conform to the
known length scales associated with real proteins). For convenience,
the three tube segments are placed inside a hard-wall spherical box of
radius 9 units -- the conformations shown are not affected by the
presence of the walls.  Our simulations were performed with standard
pivot, crank-shaft, and tail slithering move sets. Random translations
of one of the chains were also attempted. All tubes have a radius
$\Delta=1.1$. In order to minimize edge effects, the tubes were of
different lengths and the first and last points, which are not shown
in the figure, were not allowed to form any contact. One obtains 8
pairwise contacts for both the ground state arrangements shown in the
figure. Also drawn are some of the circles of radius $\Delta$ going
through several local and non-local triplets.}
\end{figure}

\newpage

\begin{figure}[h]
   \begin{center}
     \begin{math}
       \begin{array}{cc}
     (a) \includegraphics*[height=4.7cm]{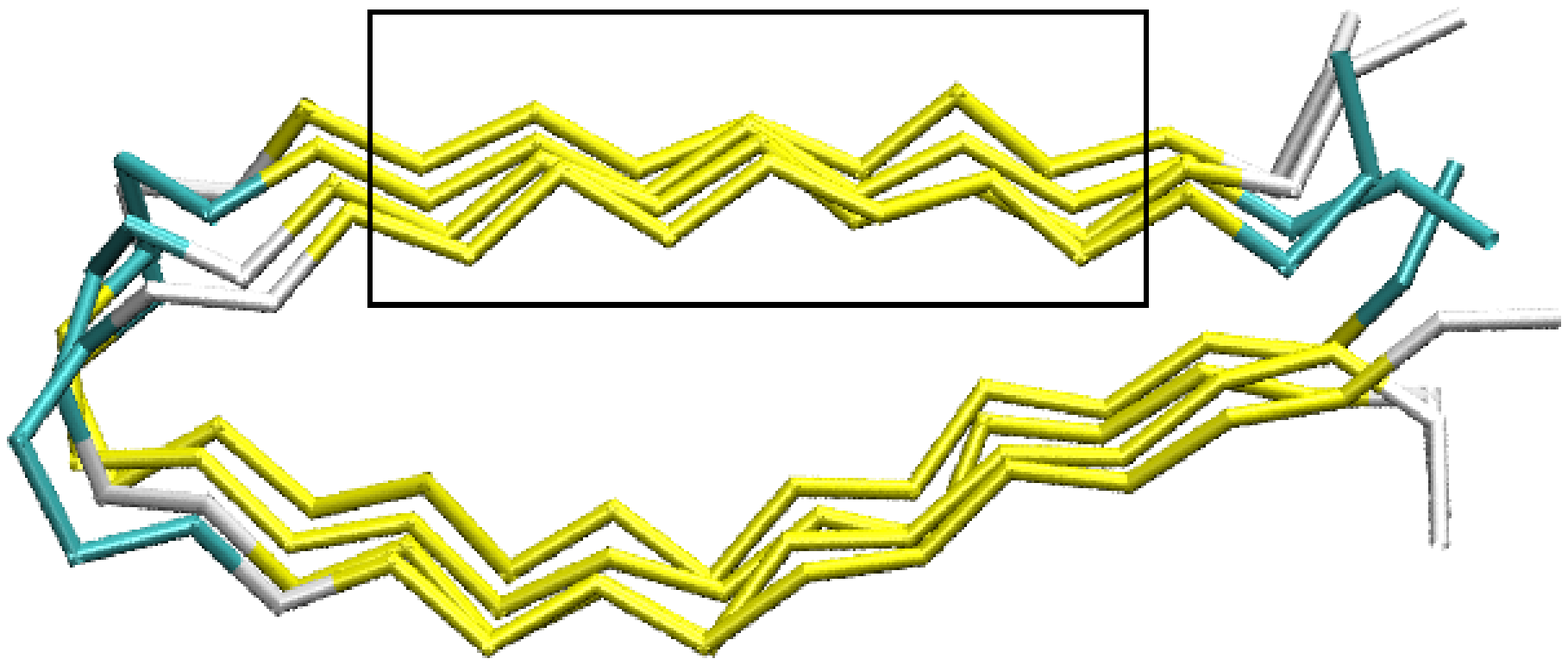}&
     (b) \includegraphics*[height=4.7cm]{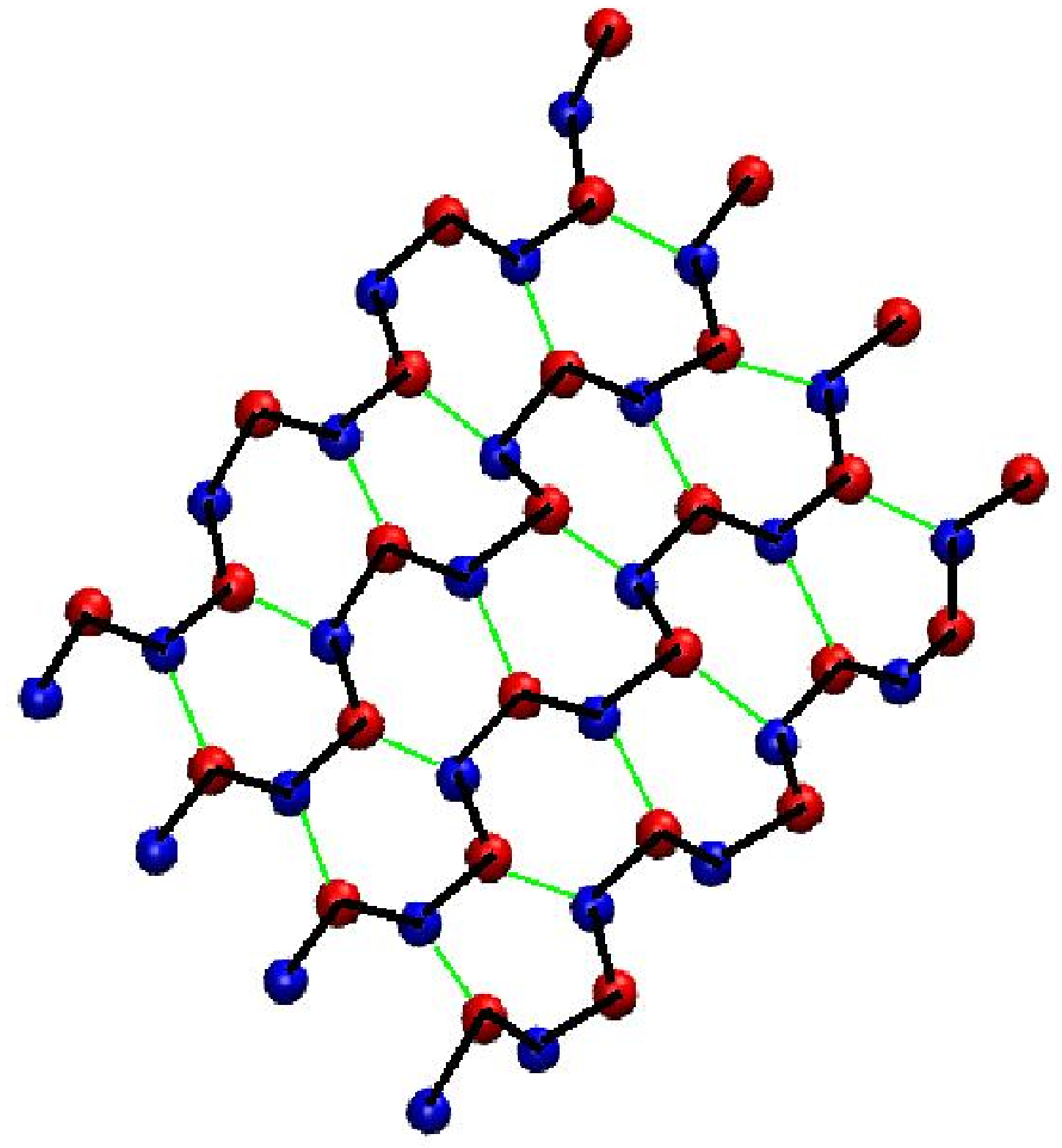}
       \end{array}
     \end{math}
   \end{center}
\caption{ Parallel $\beta$-sheets from a structural model of
CA150.WW2 protofilaments forming amyloid fibrils (Protein Data
Bank~\cite{PDB} ID code 2NNT).  The model is based on distance
constraints obtained by means of magic angle spinning (MAS) NMR
spectroscopy~\cite{Fersht06}. (a) Side view of the two $\beta$-sheets
(in yellow) forming the hairpin structure of the whole
protofilament. $C^{\alpha}$ backbone representation is
employed~\cite{VMD}. (b) Top view of the $\beta$-sheet included in the
rectangular box in (a). The representation used in (b) employs virtual
interaction centers based on main backbone atom positions.  Blue (red)
spheres are placed in the middle of the N-H (C-O) bonds and lie
approximately in the same plane. Thick black lines are drawn to
connect interaction centers along the same $\beta$-strand. Thin green
lines are drawn to represent interactions (i.e. virtual hydrogen
bonds) between neighboring strands.}
\end{figure}

\end{document}